\input harvmac

\def\im{{\rm Im}}

\def\CJ{{\cal J}}

\def\ap{\alpha'}

\def\Ga{{\Gamma}}

\def\K3{{\bf K3}}
\def\journal#1&#2(#3){\unskip, \sl #1\ \bf #2 \rm(19#3) }
\def\andjournal#1&#2(#3){\sl #1~\bf #2 \rm (19#3) }

\def\bar{\overline}
\def\hat{\widehat}

\def\tilde{\widetilde}

\def\frac#1#2{{#1\over#2}}

\def\half{\frac12}

\def\inbar{\,\vrule height1.5ex width.4pt depth0pt}
\def\IC{\relax\hbox{$\inbar\kern-.3em{\rm C}$}}
\def\IR{\relax{\rm I\kern-.18em R}}
\def\IP{\relax{\rm I\kern-.18em P}}

%
%


%
\catcode`\@=11
\def\slash#1{\mathord{\mathpalette\c@ncel{#1}}}
\overfullrule=0pt

\def\CC{{\cal C}}

\def\lam{\lambda}

\def\underrel#1\over#2{\mathrel{\mathop{\kern\z@#1}\limits_{#2}}}

\catcode`\@=12

\def\sdtimes{\mathbin{\hbox{\hskip2pt\vrule height 4.1pt depth -.3pt width
.25pt \hskip-2pt$\times$}}}

\def\det{{\rm det}}

\def\det{{\rm det}}
\def\exp{{\rm exp}}


\def \ov {\over}
\def \p {\partial}
\def \ha {{1 \ov 2}}

\def \lam {\lambda}
\def \sig {\sigma}

\def \ep {\epsilon}

\def \apr {\alpha'}
\def \m {{\rm m}}

\def\IL{\relax{\rm I\kern-.18em L}}
\def\IH{\relax{\rm I\kern-.18em H}}
\def\IR{\relax{\rm I\kern-.18em R}}
\def\IC{\relax\hbox{$\inbar\kern-.3em{\rm C}$}}
\def\IZ{{\bf Z}}
\def\CP {{\cal P }}

\def\CS{{\cal S}}





\def\makeblankbox#1#2{\hbox{\lower\dp0\vbox{\hidehrule{#1}{#2}%
   \kern -#1
   \hbox to \wd0{\hidevrule{#1}{#2}%
      \raise\ht0\vbox to #1{}
      \lower\dp0\vtop to #1{}
      \hfil\hidevrule{#2}{#1}}%
   \kern-#1\hidehrule{#2}{#1}}}%
}%
\def\hidehrule#1#2{\kern-#1\hrule height#1 depth#2 \kern-#2}%
\def\hidevrule#1#2{\kern-#1{\dimen0=#1\advance\dimen0 by #2\vrule
    width\dimen0}\kern-#2}%
\def\openbox{\ht0=1.2mm \dp0=1.2mm \wd0=2.4mm  \raise 2.75pt
\makeblankbox {.25pt} {.25pt}  }

\def\bun#1/#2{\leavevmode
   \kern.1em \raise .5ex \hbox{\the\scriptfont0 #1}%
   \kern-.1em $/$%
   \kern-.15em \lower .25ex \hbox{\the\scriptfont0 #2}%
}

\def\opensquare{\ht0=3.4mm \dp0=3.4mm \wd0=6.8mm  \raise 2.7pt
\makeblankbox {.25pt} {.25pt}  }


\def\sector#1#2{\ {\scriptstyle #1}\hskip 1mm
\mathop{\opensquare}\limits_{\lower
1mm\hbox{$\scriptstyle#2$}}\hskip 1mm}

\def\tsector#1#2{\ {\scriptstyle #1}\hskip 1mm
\mathop{\opensquare}\limits_{\lower
1mm\hbox{$\scriptstyle#2$}}^\sim\hskip 1mm}

\def\CP{{\cal P}}

\def\ap{\alpha'}

\def\IZ{{\bf Z}}

\def\ap{{\alpha'}}
\def\bx{{\bf X}}
\def\ba{{\bf A}}

\def\re{{\rm Re}}


\lref\polchinski{J. Polchinski, ``Superstring Theory'', Vol. 1,
Cambridge University Press, Cambridge, 1998.}

\lref\figueroai{J. Figueroa-O'Farrill, ``Breaking the M-waves,''
hep-th/9904124}

\lref\hkmm{J.A. Harvey, D. Kutasov, E. Martinec, and G. Moore,
``Localized Tachyons and RG Flows,''  hep-th/0111154 }

\lref\deconstruction{N. Arkami-Hamed, A.G. Cohen, D.B. Kaplan, A.
Karch, and L. Motl, ``Deconstructing $(2,0)$ and Little String
Theories'' hep-th/0110146}

\lref\atickwitten{J. Atick and E. Witten, ``The Hagedorn
transition and the number of degrees of freedom of string
theory,'' Nucl.Phys.B310:291,1988 }

\lref\ElitzurRT{ S.~Elitzur, A.~Giveon, D.~Kutasov and
E.~Rabinovici, ``From big bang to big crunch and beyond,''
arXiv:hep-th/0204189.
}

\lref\kachru{S.~Kachru and L.~McAllister, ``Bouncing Brane
Cosmologies from Warped String Compactifications'',
hep-th/0205209. }

\lref\rohm{R. Rohm, ``Spontaneous supersymmetry breaking in
supersymmetric string theories,''  Nucl.\ Phys.\ B {\bf 237}, 553
(1984). }

\lref\finiteall{G. Moore, ``Finite in All Directions,''
hep-th/9305139}

\lref\KoganNN{ I.~I.~Kogan and N.~B.~Reis, ``H-branes and chiral
strings,'' Int.\ J.\ Mod.\ Phys.\ A {\bf 16}, 4567 (2001)
[arXiv:hep-th/0107163].
}

\lref\HorowitzSR{ G.~T.~Horowitz and A.~R.~Steif, ``Strings In
Strong Gravitational Fields,'' Phys.\ Rev.\ D {\bf 42}, 1950
(1990).
}

\lref\horsteif{G.T. Horowitz and A.R. Steif, ``Is spacetime
duality violated in time dependent string solutions?'' Phys.
Lett. {\bf 250B}(1990)49}

\lref\smithpolchinski{E. Smith and J. Polchinski, ``Duality
survives time dependence,'' Phys. Lett. {\bf B} (1991) 59}

\lref\HorowitzAP{ G.~T.~Horowitz and A.~R.~Steif, ``Singular
String Solutions With Nonsingular Initial Data,'' Phys.\ Lett.\ B
{\bf 258}, 91 (1991).
}

\lref\SusskindIF{ L.~Susskind, L.~Thorlacius and J.~Uglum, ``The
Stretched horizon and black hole complementarity,'' Phys.\ Rev.\
D {\bf 48}, 3743 (1993) [arXiv:hep-th/9306069].
}

\lref\GibbonsJB{ G.~W.~Gibbons, ``Quantized Fields Propagating In
Plane Wave Space-Times,'' Commun.\ Math.\ Phys.\  {\bf 45}, 191
(1975).
}

\lref\BirrellIX{ N.~D.~Birrell and P.~C.~Davies, ``Quantum Fields
In Curved Space,'' {\it  Cambridge, Uk: Univ. Pr. ( 1982) 340p}. }

\lref\VenezianoPZ{ G.~Veneziano, ``String cosmology: The pre-big
bang scenario,'' arXiv:hep-th/0002094.
}

\lref\KhouryBZ{ J.~Khoury, B.~A.~Ovrut, N.~Seiberg,
P.~J.~Steinhardt and N.~Turok, ``From big crunch to big bang,''
arXiv:hep-th/0108187;
N.~Seiberg, ``From big crunch to big bang - is it possible?,''
arXiv:hep-th/0201039.
}

\lref\RussoIK{
J.~G.~Russo and A.~A.~Tseytlin,
``Magnetic flux tube models in superstring theory,''
Nucl.\ Phys.\ B {\bf 461}, 131 (1996)
[arXiv:hep-th/9508068].
}

\lref\BanadosWN{ M.~Banados, C.~Teitelboim and J.~Zanelli, ``The
Black Hole In Three-Dimensional Space-Time,'' Phys.\ Rev.\ Lett.\
{\bf 69}, 1849 (1992) [arXiv:hep-th/9204099].
}
\lref\BanadosGQ{ M.~Banados, M.~Henneaux, C.~Teitelboim and
J.~Zanelli, ``Geometry of the (2+1) black hole,'' Phys.\ Rev.\ D
{\bf 48}, 1506 (1993) [arXiv:gr-qc/9302012].
}
\lref\ShiraishiHF{ K.~Shiraishi and T.~Maki, ``Quantum fluctuation
of stress tensor and black holes in three-dimensions,'' Phys.\
Rev.\ D {\bf 49}, 5286 (1994).
}

\lref\SteifZV{ A.~R.~Steif, ``The Quantum Stress Tensor In The
Three-Dimensional Black Hole,'' Phys.\ Rev.\ D {\bf 49}, 585
(1994) [arXiv:gr-qc/9308032].
}

\lref\LifschytzEB{ G.~Lifschytz and M.~Ortiz, ``Scalar Field
Quantization On The (2+1)-Dimensional Black Hole Background,''
Phys.\ Rev.\ D {\bf 49}, 1929 (1994) [arXiv:gr-qc/9310008].
}

\lref\PolchinskiMF{ J.~Polchinski, ``Critical Behavior Of Random
Surfaces In One-Dimension,'' Nucl.\ Phys.\ B {\bf 346}, 253
(1990).
}

\lref\PolchinskiUQ{ J.~Polchinski, ``Classical Limit Of
(1+1)-Dimensional String Theory,'' Nucl.\ Phys.\ B {\bf 362}, 125
(1991).
}

\lref\DiFrancescoSS{ P.~Di Francesco and D.~Kutasov,
``Correlation functions in 2-D string theory,'' Phys.\ Lett.\ B
{\bf 261}, 385 (1991).
}

\lref\DiFrancescoUD{ P.~Di Francesco and D.~Kutasov, ``World
sheet and space-time physics in two-dimensional (Super)string
theory,'' Nucl.\ Phys.\ B {\bf 375}, 119 (1992)
[arXiv:hep-th/9109005].
}

\lref\MooreGB{ G.~W.~Moore and R.~Plesser, ``Classical scattering
in (1+1)-dimensional string theory,'' Phys.\ Rev.\ D {\bf 46},
1730 (1992) [arXiv:hep-th/9203060].
}

\lref\KimMI{ H.~Kim, J.~S.~Oh and C.~R.~Ahn, ``Quantisation of
conformal fields in AdS(3) black hole spacetime,'' Int.\ J.\ Mod.\
Phys.\ A {\bf 14}, 2431 (1999) [arXiv:hep-th/9708072].
}
\lref\CarlipQV{ S.~Carlip, ``The (2+1)-Dimensional black hole,''
Class.\ Quant.\ Grav.\  {\bf 12}, 2853 (1995)
[arXiv:gr-qc/9506079].
}
\lref\MaldacenaKR{ J.~M.~Maldacena, ``Eternal black holes in
Anti-de-Sitter,'' arXiv:hep-th/0106112.
} \lref\HorowitzAP{ G.~T.~Horowitz and A.~R.~Steif, ``Singular
String Solutions With Nonsingular Initial Data,'' Phys.\ Lett.\ B
{\bf 258}, 91 (1991).
}
\lref\BalasubramanianRY{ V.~Balasubramanian, S.~F.~Hassan,
E.~Keski-Vakkuri and A.~Naqvi, ``A space-time orbifold: A toy
model for a cosmological singularity,'' arXiv:hep-th/0202187.
}
\lref\GutperleAI{ M.~Gutperle and A.~Strominger, ``Spacelike
branes,'' arXiv:hep-th/0202210.
}
\lref\CornalbaFI{ L.~Cornalba and M.~S.~Costa, ``A New
Cosmological Scenario in String Theory,'' arXiv:hep-th/0203031.
}
\lref\NekrasovKF{ N.~A.~Nekrasov, ``Milne universe, tachyons, and
quantum group,'' hep-th/0203112.
}

\lref\SenNU{ A.~Sen, ``Rolling Tachyon,'' arXiv:hep-th/0203211.
}

\lref\HiscockJQ{ W.~A.~Hiscock, ``Quantized fields and chronology
protection,'' arXiv:gr-qc/0009061.
}

\lref\HawkingNK{ S.~W.~Hawking, ``The Chronology protection
conjecture,'' Phys.\ Rev.\ D {\bf 46}, 603 (1992).
}

\lref\HawkingPK{ S.~W.~Hawking, ``The Chronology Protection
Conjecture,'' {\it Prepared for 6th Marcel Grossmann Meeting on
General Relativity (MG6), Kyoto, Japan, 23-29 Jun 1991}. }

\lref\BrodskyDE{ S.~J.~Brodsky, H.~C.~Pauli and S.~S.~Pinsky,
``Quantum chromodynamics and other field theories on the light
cone,'' Phys.\ Rept.\  {\bf 301}, 299 (1998)
[arXiv:hep-ph/9705477].
}

\lref\SimonMA{ J.~Simon, ``The geometry of null rotation
identifications,'' arXiv:hep-th/0203201.
}

\lref\fluxbranes{J.~Figueroa-O'Farrill and J.~Simon, ``Generalized
supersymmetric fluxbranes,'' JHEP {\bf 0112}, 011 (2001)
[arXiv:hep-th/0110170].
}

\lref\lawrence{A.~Lawrence, ``On the instability of 3d null
singularities'', hep-th/0205288.}

\lref\hp{G.~Horowitz and J.~Polchinski, to appear.}

\lref\mwaver{JM.~Figueroa-O'Farrill, ``Breaking the M-waves,''
Class.\ Quant.\ Grav.\ {\bf 17}, 2925 (2000)
[arXiv:hep-th/9904124].}

\lref\longpaper{H. Liu, G. Moore, and N. Seiberg, To appear.}

\lref\polchinski{J. Polchinski, ``Superstring Theory'', Vol. 1,
Cambridge University Press, Cambridge, 1998.}

\lref\BanadosWN{ M.~Banados, C.~Teitelboim and J.~Zanelli, ``The
Black Hole In Three-Dimensional Space-Time,'' Phys.\ Rev.\ Lett.\
{\bf 69}, 1849 (1992) [arXiv:hep-th/9204099].
}
\lref\BanadosGQ{ M.~Banados, M.~Henneaux, C.~Teitelboim and
J.~Zanelli, ``Geometry of the (2+1) black hole,'' Phys.\ Rev.\ D
{\bf 48}, 1506 (1993) [arXiv:gr-qc/9302012].
}

\lref\hkmm{J.A. Harvey, D. Kutasov, E. Martinec, and G. Moore,
``Localized Tachyons and RG Flows,''  hep-th/0111154 }

\lref\deconstruction{N. Arkami-Hamed, A.G. Cohen, D.B. Kaplan, A.
Karch, and L. Motl, ``Deconstructing $(2,0)$ and Little String
Theories'' hep-th/0110146}

\lref\atickwitten{J. Atick and E. Witten, ``The Hagedorn
transition and the number of degrees of freedom of string
theory,'' Nucl.Phys.B310:291-334,1988 }

\lref\LMS{ H.~Liu, G.~Moore and N.~Seiberg, ``Strings in a
time-dependent orbifold,'' arXiv:hep-th/0204168.
}

\lref\BalasubramanianAM{ V.~Balasubramanian and S.~F.~Ross, ``The
dual of nothing,'' arXiv:hep-th/0205290.
}

\lref\MooreZC{ G.~W.~Moore, ``Finite In All Directions,''
arXiv:hep-th/9305139.
}

\lref\horsteif{G.T. Horowitz and A.R. Steif, ``Is spacetime
duality violated in time dependent string solutions?'' Phys.
Lett. {\bf 250B}(1990)49}

\lref\smithpolchinski{E. Smith and J. Polchinski, ``Duality
survives time dependence,'' Phys. Lett. B {\bf 263} 59 (1991).}

\lref\fabmcg{M.~Fabinger and J.~McGreevy, to appear.}

\lref\HorowitzAP{ G.~T.~Horowitz and A.~R.~Steif, ``Singular
String Solutions With Nonsingular Initial Data,'' Phys.\ Lett.\ B
{\bf 258}, 91 (1991).
}

\lref\ArkaniHamedIE{ N.~Arkani-Hamed, A.~G.~Cohen, D.~B.~Kaplan,
A.~Karch and L.~Motl, ``Deconstructing (2,0) and little string
theories,'' arXiv:hep-th/0110146.
}

\lref\tseytlinup{ A.~A.~Tseytlin, ``Exact string solutions and
duality,'' arXiv:hep-th/9407099;
C.~Klimcik and A.~A.~Tseytlin, unpublished (1994); A. A.
Tseytlin, unpublished (2001).}

\lref\aharony{ O.~Aharony, M.~Fabinger, G.~Horowitz, and
E.~Silverstein, ``Clean Time-Dependent String Backgrounds from
Bubble Baths,'' arXiv:hep-th/0204158.}

\lref\CornalbaNV{ L.~Cornalba, M.~S.~Costa and C.~Kounnas, ``A
resolution of the cosmological singularity with orientifolds,''
arXiv:hep-th/0204261.
}

\lref\CrapsII{ B.~Craps, D.~Kutasov and G.~Rajesh, ``String
propagation in the presence of cosmological singularities,''
arXiv:hep-th/0205101.
}

\lref\TolleyCV{ A.~J.~Tolley and N.~Turok, ``Quantum fields in a
big crunch / big bang spacetime,'' arXiv:hep-th/0204091.
}

\lref\reedsimon{M. Reed and B. Simon, ``Methods of Mathematical
Physics  Vol. 3: Scattering Theory,'' {\it  New York, Usa:
Academic ( 1979) 463p}.}

\lref\manogue{C. Manogue and J. Schray, ``Finite Lorentz Transformations,
Automorphisms, and Division Algebras,'' hep-th/9302044}

\lref\TakayanagiJJ{ T.~Takayanagi and T.~Uesugi, ``Orbifolds as
Melvin geometry,'' JHEP {\bf 0112}, 004 (2001)
[arXiv:hep-th/0110099].
}

\lref\TakayanagiAJ{ T.~Takayanagi and T.~Uesugi, ``D-branes in
Melvin background,'' JHEP {\bf 0111}, 036 (2001)
[arXiv:hep-th/0110200].
}

\lref\KiritsisKZ{ E.~Kiritsis and B.~Pioline, ``Strings in
homogeneous gravitational waves and null holography,''
arXiv:hep-th/0204004.
}

\lref\blau{M. Blau, J. Figueroa-O'Farrill, and G. Papadopoulos,
``Penrose limits, supergravity, and brane dynamics,''
hep-th/0202111}

\lref\WittenKN{
E.~Witten,
``Quantum gravity in de Sitter space,''
arXiv:hep-th/0106109.
}

\lref\BanksFE{
T.~Banks,
``Cosmological breaking of supersymmetry or little Lambda goes
back to  the future. II,''
arXiv:hep-th/0007146.
}

\lref\soldate{I.J. Muzinich and M. Solate,
``High-energy unitarity of gravitation and strings,''
Phys. Rev. {\bf D37}(1988) 359}

\lref\amati{
D.~Amati, M.~Ciafaloni and G.~Veneziano,
``Superstring Collisions At Planckian Energies,''
Phys.\ Lett.\ B {\bf 197}, 81 (1987);
D.~Amati, M.~Ciafaloni and G.~Veneziano,
``Classical And Quantum Gravity Effects From Planckian Energy
Superstring Collisions,''
Int.\ J.\ Mod.\ Phys.\ A {\bf 3}, 1615 (1988).
}

\lref\robbins{D.~Robbins and S.~Sethi, to appear.}


\Title{\vbox{\baselineskip12pt \hbox{hep-th/0206182}
\hbox{RUNHETC-2002-19}
\hbox{NI02014-MTH}
}}%
{\vbox{\centerline{Strings in Time-Dependent Orbifolds} }}

\smallskip
\centerline{Hong Liu, Gregory Moore}
\medskip

\centerline{\it Department of Physics, Rutgers University}
\centerline{\it Piscataway, New Jersey, 08855-0849}

\bigskip

\centerline{and}

\bigskip

\centerline{Nathan Seiberg}
\medskip
\centerline{\it School of Natural Sciences}
\centerline{\it Institute for Advanced Study}
\centerline{\it Einstein Drive,Princeton, NJ 08540}

\smallskip

\vglue .3cm

\bigskip
\noindent We continue and extend our earlier investigation
``Strings in a Time-Dependent Orbifold'' (hep-th/0204168).  We
formulate conditions for an orbifold to be amenable to
perturbative string analysis and classify the low dimensional
orbifolds satisfying these conditions. We analyze the tree and
torus amplitudes of some of these orbifolds. The tree amplitudes
exhibit a new kind of infrared divergences which are a result of
some ultraviolet effects.  These UV enhanced IR divergences can
be interpreted as due to back reaction of the geometry. We argue
that for this reason the three dimensional parabolic orbifold is
not amenable to perturbation theory. Similarly, the smooth four
dimensional null-brane tensored with sufficiently few noncompact
dimensions also appears problematic. However, when the number of
noncompact dimensions is sufficiently large perturbation theory
in these time dependent backgrounds seems consistent.

\Date{June 20, 2002}



\newsec{Introduction}

Time dependent physics poses a large number of interesting
conceptual and technical problems in quantum field theory, quantum
gravity, and in string theory. So far  very little work has been
done on time-dependent backgrounds in  the framework of string
theory, although the situation is beginning to change (see, e.g.\
\refs{\HorowitzAP\HorowitzSR\figueroai\fluxbranes\KhouryBZ\BalasubramanianRY
\CornalbaFI\NekrasovKF\KiritsisKZ\TolleyCV\SimonMA\GutperleAI\SenNU
\aharony\LMS\ElitzurRT\CornalbaNV\CrapsII\BalasubramanianAM
\kachru\lawrence-\hp} for recent discussions of this issue). Here,
we continue the investigation of strings in time dependent
orbifolds which we started in \LMS. We are adopting a conservative
approach, simply trying to follow the standard rules of
perturbative string theory, and trying to imitate the
constructions and methods which have been proven useful in time
independent setups.

In order to be able to apply perturbative string methods we have
to deal with the following issues: \item{1.} We start with a
solution of the classical equations of motion; i.e.\ a worldsheet
conformal field theory. This conformal field theory is most
tractable when it is based on an orbifold of flat spacetime.
Therefore, we will mod out $\IR^{1,n}$ by a subgroup $\Gamma$ of
its symmetry group which is the Poincare group in $n+1$ dimensions
$ \CP(1,n)$.  We will denote the orbifold by
$\CO=\IR^{1,n}/\Gamma$.  In order to have nontrivial time
dependence $\Gamma$ should not be a subgroup of the Euclidean
group in $n$ dimensions as in ordinary Euclidean orbifolds.
\item{2.}  In Euclidean orbifolds the singularities of the target
space are at fixed points of $\Gamma$.  Lorentzian orbifolds
typically have also other potential problems. The quotient by
$\Gamma$ can make time non-orientable.  It can also create closed
null curves (CNC's) or closed time-like curves (CTC).  These are
potentially problematic because they can lead to divergent
expectation values of composite operators like the stress tensor
and to large back reaction of the metric. We examine
orbifolds which are time-orientable and have no CTC's,
although we will
allow singularities, non-Hausdorff spaces, and the possibility
of CNC's. We will also see examples in which the orbifold has no CNC's
but has closed curves whose invariant length is arbitrarily small;
we will examine the consequences of these curves.
\item{3.} In
order to help ensure the stability of the background against
radiative corrections we will look for orbifolds which leave
unbroken some amount of supersymmetry.  This supersymmetry can
guarantee that some of the back reaction due to CNC when they are
present is harmless.  Having such supersymmetry leads to a null
Killing vector $\partial_{x^-}$ (see \refs{\figueroai,\fluxbranes}
and section 7 below).  We can therefore use lightcone frame and
treat $x^+$ as time.  The orbifold $\CO$ is thus foliated by
spaces $\CF_{x^+}$ of fixed $x^+$.  The lightcone description also
guarantees that the vacuum of the second quantized theory is
trivial, and there is no particle production in the system
\GibbonsJB. \item{4.} Even if supersymmetry guarantees that the
zero point function and the one point functions vanish to all
orders, the question of back reaction is far from obvious.  Vertex
operators correspond to small deformations of the background.  If
they are singular  at some point in spacetime, they can lead to
infinite energy density there and to large change in the
gravitational field.  Such an effect can render perturbation
theory invalid. For smooth backgrounds vertex operators which are
singular at some point in spacetime can be excluded, but in
various singular orbifolds all vertex operators are singular.  We
will discuss this issue in more detail below in some examples.  We
will also show that sometimes a smooth background with smooth
vertex operators can also suffer from large back reaction. In
particular, we will see that the back reaction is reflected in
``UV enhanced IR divergences'', which may also signal that one
needs to consider more subtle asymptotic states.

In section 2 we will examine two orbifolds.  The first, which we
will refer to as the parabolic orbifold (it is also called the
``null orbifold''), was introduced in \HorowitzAP\ and briefly
studied in \tseytlinup. Its geometry was further explored
recently in \SimonMA\ and strings in this background were
investigated in \LMS. The second orbifold, the ``null-brane'' was
described in \refs{\fluxbranes,\SimonMA}.  In sections 3 -- 6 we
will slightly extend the analysis of string theory in the
background of the parabolic orbifold in \LMS, and will apply the
same techniques to the case of the null-brane background.  In
section 3 we consider the functions on the orbifolds which are
important when constructing vertex operators.  In section 4 we
consider the canonical quantization of free strings.  In section
5 we analyze the torus partition function, and in section 6 we
consider tree level S-matrix amplitudes.  In section 7 we comment on
the general classification of such models and describe several
new orbifolds which exhibit new phenomena.

Strings in the null-brane orbifold are also studied by D. Robbins
and S. Sethi \robbins\ and by M.~Fabinger and J.~McGreevy
\fabmcg. Also, A.~Lawrence \lawrence\ and G.~Horowitz and
J.~Polchinski \hp\ reached conclusions related to ours about the
back reaction and the validity of perturbation theory in the
parabolic orbifold and the null-brane.

\newsec{The parabolic orbifold and the null-brane}

\subsec{Coordinates and Metric}

In \LMS\ we analyzed in detail string theory on a parabolic
orbifold, with target space $\bigl(\IR^{1,2}/\Gamma \bigr) \times
\CC^{\perp}$ where $\Gamma\cong \IZ$ is a parabolic subgroup of
the three dimensional Lorentz group ${\rm Spin}(1,2) \cong
SL(2,R)$. $\CC^{\perp}$ is a transverse Euclidean CFT which is
invariant under the action of $\Gamma$. More explicitly, writing
the Lorentzian metric for $\IR^{1,2}$ as $ds^2=-2dx^+dx^-+dx^2$,
the generator of the orbifold group $\Ga$ acts as
 \eqn\partrans{ X=
 \pmatrix{ x^+ \cr x^{~}\cr x^-\cr} \quad \rightarrow\quad
 g_0\cdot X=e^{ 2 \pi \CJ} X = \pmatrix{ x^+ \cr x + 2 \pi  x^+\cr
 x^- + 2\pi x + \half (2 \pi)^2  x^+ \cr} ;\qquad
 \CJ=\pmatrix{0&0&0\cr
 1&0&0\cr0&1&0}
 }
That is, $ g_0= \exp\bigl(2 \pi i J\bigr)$ where we take the Lie
algebra generator
 \eqn\liegen{ J = J^{+x} = {1\over \sqrt{2}}(J^{0x} + J^{1x} )}
corresponding to a linear combination of a boost and a rotation.

For some purposes it is convenient to use coordinates \foot{This
is a special case of the transformation from ``Brinkman'' to
``Rosen'' coordinates in the theory of pp waves \blau.}
 \eqn\newcoordii{ \eqalign{
 x^+ &= y^+ \cr
 x & = y^+ y \cr
 x^- & = y^-  + \ha y^+y^2 \cr} }
in terms of which the metric becomes
 \eqn\ymetric{ds^2=-2dy^+dy^-+(y^+)^2dy^2}
and the orbifold identification is simply $y \sim y +  2\pi n$.
The spacetime \ymetric\ may be visualized as two cones
(parametrized  by $y^+$ and $y$) with a common tip at $y^+ =0$,
crossed with the real line (for $y^-$).  This description is not
valid at the singularity $x^+=y^+=0$ where this coordinate system
is not valid.

The parabolic orbifold has a few attractive features which fit
nicely with the criteria outlined in the introduction: \item{1.}
The orbifold has a null isometry which allows us to use light-cone
evolution. There is no string or particle production in the second
quantized theory. \item{2.} The orbifold has a covariant spinor
and thus preserves half of the supersymmetries in the superstring
theory. As a result, one can show that one-loop cosmological
constant and the massless tadpoles vanish. \item{3.} There are
{\it no} closed time-like curves, while there are closed null
curves only in the hyperplane $x^+ =0$. \item{4.} In terms of
light-cone time, the orbifold describes the big crunch of a circle
at $x^+ =0$, followed by a big bang. There is a null singular line
at $x^+=0$. This provides an interesting toy model for
understanding the big crunch/big bang singularity in cosmology. It
is also related to the singularity of the massless BTZ black hole
\refs{\BanadosWN,\BanadosGQ}. \item{5.} If the spacetime does not
end at the singularity, one might be able to define an $S$-matrix
as a natural observable.

In \LMS\ we observed that there are divergences in the $S$-matrix
of the parabolic orbifold for special kinematic configurations.
We suggested there that they might signal the breakdown of
perturbation theory.  Below we will analyze these divergences in
more detail, and will argue that this conclusion is unavoidable.
Therefore, even if such an S-matrix exists, it cannot be computed
in perturbation  theory.

In this paper we consider a few generalizations of the parabolic
orbifold by allowing $\Ga$ to act nontrivially on the transverse
spacetime $\CC^{\perp}$  or by including more generators, in a way
that the above attractive features remain.  In the following we
will discuss in detail a  simple family of orbifolds of which the
parabolic orbifold is a special example. In particular we will
show that this more general family does allow a perturbative
computation of the S-matrix. Other generalizations and
classifications will be discussed in sec. 7.

We start with flat four dimensional spacetime $\IR^{1,3}$ with
the metric
 \eqn\xmetrici{ds^2=-2dx^+dx^-+dx^2 + dz^2 \ .}
and consider the orbifolds obtained by identifying
 \eqn\xidentii{
 X \sim e^{2 \pi n \CJ} X, \qquad z \sim z + 2 \pi n R, \qquad n
 \in \IZ \ }
where the column vector $X$ and matrix $\CJ$ are the same as in
\partrans. The generator of the identification is now given by
 \eqn\orbgrn{g_0 = e^{2 \pi i ( J +  R  p_z)}, }
where $J$ is given by \liegen\ and $p_z$ is the generator for
translation in $z$ direction.

The above orbifold, called the null-brane, was introduced in
\refs{\fluxbranes,\SimonMA}. The null-brane has a continuous
modulus $R$ and the parabolic orbifold corresponds to the
singular limit $R \to 0$. The geodesic distance  between a point
and its $n$'th image is $2 \pi n \sqrt{R^2 + (x^+)^2}$.  At finite
$R$ the spacetime is regular and Hausdorff everywhere with no
closed causal curves. Thus it may also be considered as a
regularization of the parabolic orbifold.

The orbifold action breaks the Poincar\'e symmetry of
$\IR^{1,3}$, leaving only three of its Lie algebra generators
unbroken.  These are $J$, $p^+=-p_-$ which shifts $x^-$ by a
constant, and $p_z$ which generates translations in the
$z$-direction.  The null Killing vector associated with $p^+$
allows us to use a light-cone evolution treating $x^+$ as time.
Thus again there is no particle production in the second
quantized theory. As in the parabolic orbifold,  for an
appropriate choice of the sign of the generator \orbgrn\ acting
on spinors, the group $\Gamma$  leaves one half of all spinor
components  invariant. When the superstring is compactified on a
null-brane it preserves half of the supercharges.  These
supercharges square to the Killing vector $p^+$.

Let us now look at the geometry of the null-brane in more detail.
We will change to two different coordinate systems. First, for
$x^+ \neq 0$ it is convenient to define $y^\pm,y$ as in
\newcoordii\ and to express the metric \xmetrici\ as
 \eqn\ymetrici{ds^2=-2dy^+dy^-+(y^+)^2dy^2  + dz^2}
and the orbifold identification \xidentii\ becomes
 \eqn\yidentii{\pmatrix{y^+\cr y\cr y^- \cr z} \sim
 \pmatrix{y^+\cr
 y + 2\pi n\cr y^- \cr z + 2 \pi n R}}
If we also define
 \eqn\zud{z = R y + u}
where $u$ is a noncompact coordinate, then the metric becomes
 \eqn\fmetric{\eqalign{
 ds^2  & = - 2 dy^+dy^- + du^2 + \left(R^2  + (y^+)^2
 \right) d y^2 + 2R  du \, d y \cr
 & =  - 2 dy^+dy^- + {(y^+)^2 \ov R^2 + (y^+)^2} d u^2
 + ( R^2 + (y^+)^2) \left(dy + {R \ov R^2 + (y^+)^2} du \right)^2
 }}
\fmetric\ represents a circle fibration over a three-dimensional
manifold parameterized by $(y^+, y^-,u)$, where the the circle
has a radius $ \sqrt{R^2 + (y^+)^2}$ and the fibration has a
connection with
 \eqn\conne{
 F = -{R y^+ \ov (R^2 + (y^+)^2)^2} d y^+ \wedge du \ .
 }
It is manifest in the first line of \fmetric\ that as $R \to 0$
we get the metric of the parabolic orbifold in $y$ coordinate
times a transverse noncompact direction
 \eqn\parabolm{
 ds^2 = - 2 dy^+ dy^- + (y^+) dy^2 + du^2  \ .
 }
Note that since the coordinate transformation \newcoordii\ is
singular at $y^+ =0$, the $y^+ \to 0$ limit of \fmetric\ should
be treated with caution.

Another set of global coordinates \fluxbranes\ can be obtained by
taking
 \eqn\gloco{
z = R \theta, \qquad
 \tilde X = e^{- {z \ov R} \CJ} X, \qquad
 }
i.e. explicitly,
 \eqn\expcor{ \eqalign{
 x^+ &= \tilde x^+ \cr
 x & = \tilde x + \theta \tilde x^+ \cr
 x^- & = \tilde x^-  + \theta \tilde x + \ha \theta^2 \tilde
 x^+ \cr} }
In these coordinates the metric becomes
 \eqn\timetr{\eqalign{
 ds^2 & = - 2 d \tilde x^+ d \tilde x^- + d \tilde x^2
 + \left(R^2 + (\tilde x^+)^2 \right) d \theta^2
 + 2 d \theta (\tilde x^+ d \tilde x - \tilde x d \tilde x^+ ) \cr
 & = - 2 d \tilde x^+ d \tilde x^- + d \tilde x^2
 - {1 \ov R^2 + (\tilde x^+)^2} (\tilde x^+ d \tilde x
 - \tilde x d \tilde x^+ )^2 \cr
 & \qquad \qquad
 + \left(R^2 + (\tilde x^+)^2 \right)
 \left(d \theta + {1 \ov  R^2 + (\tilde x^+)^2} (\tilde x^+ d
 \tilde x
 - \tilde x d \tilde x^+ )\right)^2
 }}
Note that the above metric has a constant determinant $-R^2$.
\timetr\ represents a circle fibration over a 3-manifold
parametrized by $\tilde X$. The latter has nonsingular metric
with $\det g = {R^2\over R^2 + (\tilde x^+)^2}$. Moreover there is
a connection on the circle fibration  with
 \eqn\fluxbrane{
 F = {2 R^2\over (R^2 + (\tilde x^+)^2)^2} d \tilde x^+ \wedge d
 \tilde x\ .}
There is a strong formal similarity between this solution and the
standard Melvin universe. As in the parabolic orbifold, $\tilde
x^+$ plays a dual role of a time and a ``radial variable.'' Note
that in the present case it is clear that $-\infty< \tilde x^+ <
+\infty$, so that we have the ``two-cone'' theory. It would be of
some interest to extend the present discussion to the analog of
the two-parameter Melvin solutions
\refs{\RussoIK\TakayanagiJJ-\TakayanagiAJ} with nontrivial
$H$-flux but we will not attempt this in the present paper.

The coordinate system $(y^\mu, z)$ is not geodesically complete.
Its completion is given by the system $(\tilde X, \theta)$
related by
\eqn\ytoxtild{
\eqalign{
\tilde x^+ & = y^+ \cr
\tilde x & = y^+(y-z/R)\cr
\tilde x^- & = y^- + \half y^+(y- z/R)^2\cr
\theta & = z/R\cr}
}
The $R \to 0$ limit is somewhat subtle in these coordinates since
the $\theta$ coordinate used in \gloco\ and \expcor\ is not well
defined in this limit.

In terms of light-cone time $x^+$, the null-brane describes an
infinite size circle in the far past shrinking to a minimal radius
$R$ at $x^+=0$ and then expanding to infinite size in the remote
future. An interesting feature of the causal structure of the
parabolic orbifold is that every point $P$ with $y^+ > 0$ is
always in the causal future of the every point $\tilde P$ with
$y^+ < 0$. In the null-brane geometry, while this is no longer so,
there is a close analogue. The geodesic distance square between a
point $\CP_1$ and the $n^{th}$ image of a point $\CP_2$ is
 \eqn\newdetn{
 \Delta_n(\CP_1,\CP_2) = -2 \Delta y^+ \Delta y^- + y^+_1 y_2^+
 (\Delta y - 2\pi n)^2 +  (\Delta z - 2\pi n R)^2 }
The large $n$ behavior of the above equation for fixed $\CP_1,
\CP_2$ is
 \eqn\largen{
 \Delta_n(\CP_1,\CP_2)  \sim (y_1^+ y_2^+ + R^2) (2\pi n)^2
 }
Thus, points with $y_1^+ y_2^+ + R^2 < 0$ are necessarily
timelike separated.

\newsec{Solutions of the wave equation on the null-brane}

In \LMS\ we discussed the solutions of the wave equation on the
parabolic orbifold. Here we give an analogous treatment for the
null-brane. These functions, which are easily derived from those
of the parabolic orbifold, are important for studying first
quantized particles on this spacetime and for constructing the
vertex operators in string theory.

We start by considering functions on the covering space
$\IR^{1,3}$. We diagonalize the Killing vectors $J, p^+,p_z$ and
the Laplacian :
  \eqn\waveeq{\eqalign{
 &  \left[-2 {\p \over \p x^+}{\p \over \p x^-}+ {\p^2
 \over \p x^2} + {\p^2  \over \p z^2} \right] \psi = m^2 \psi \cr
 & \hat J \psi  = -i \bigl( x^+{\p
 \over \p x} + x {\p \over \p x^-} \bigr) \psi = J \psi \cr
 & \hat p^+ \psi = i {\p \ov \p x^-} \psi = p^+ \psi, \qquad
 \hat p_z  \psi = - i {\p \ov \p z}  \psi = k \psi  \ .
 }}
Following the discussion in \LMS, the eigenfunctions functions
can be written as
 \eqn\invw{\eqalign{
 \psi_{p^+,J,k,m^2} & = \sqrt{1\over i  x^+}\
 \exp \left[-ip^+x^--i{m^2 + k^2 \over 2p^+} x^+ + i{ p^+
 \over 2 x^+}(x-\xi)^2  + ikz\right] \cr
  & = \int_{-\infty}^\infty {dp  \over \sqrt{2 \pi p^+} }\
 e^{-ip\xi} \; \exp \left[-i p^+ x^- - i {p^2 + k^2 + m^2 \ov 2
 p^+} x^+ + i px +ikz  \right]
 }}
with
 \eqn\xzerodef{\xi=-{J \over p^+} \ .}
The second line of \invw\ shows that $J$-eigenfunctions can be
obtained from the standard momentum eigenfunctions by a Fourier
transform. The functions $\psi_{p^+,J,k,m^2} $ form a complete
basis of functions for fixed $x^+$ with the inner product
 \eqn\naomzcx{
 \int_{-\infty}^{\infty} dx^- dx dz\,
 \psi^*_{p^+,J,k,m^2}  \,
 \psi_{\tilde p^+, \tilde J, \tilde k, m^2}
 = (2 \pi)^3 \delta (p^+ - \tilde p^+) \, \delta (J-\tilde J) \,
 \delta(k - \tilde k) \ .}

Consider now the null-brane orbifold $\CO=\IR^{1,3}/\Gamma$. Under
the identification \xidentii\ the function \invw\ transforms as
 \eqn\trab{\psi_{p^+,J,k,m^2}  \to
 e^{2 \pi i n (J +  k R)}\,\psi_{p^+,J,k,m^2}  }
It then follows that in the null-brane functions should satisfy
the quantization condition
 \eqn\incn{
 J + k R = n \in \IZ \ .}
Therefore, we label the basis functions
 \eqn\vdef{V_{p^+,J,n,m^2}=\psi_{p^+,J,k={n-J\over R},m^2}}
The orbifold $\CO$ is foliated by equal $x^+$ spaces $\CF_{x^+}$
and the complete basis functions $V_{p^+,J,n,m^2}$ satisfy
 \eqn\vinner{\int_{\CF_{x^+}}V^*_{p^+,J,n,m^2}  \,
 V_{\tilde p^+, \tilde J, \tilde n, m^2}
 = (2 \pi)^3 R \, \delta_{n,\tilde n} \, \delta (p^+ - \tilde p^+)
 \, \delta (J-\tilde J) \,  \ .}

In terms of the coordinate system \newcoordii\zud\ the functions
\vdef\ can be written as
 \eqn\pshffun{
 V_{p^+,J,n,m^2} = \sqrt{1 \over i y^+}\ \exp
 \left[-ip^+y^-+ i n y +i{J^2 \over 2p^+y^+}
 -i{\left({n-J\over R}\right)^2 + m^2 \over 2p^+} y^+  +i\left(
 {n-J\over R}\right)u\right] }
and in terms of the coordinates \gloco\ they are
 \eqn\glofunc{V_{p^+,J,n,m^2} = \sqrt {1\over i  \tilde x^+}
 \exp \left[-ip^+ \tilde x^--i{\left({n-J\over R}\right)^2 +m^2
 \over 2p^+} \tilde x^+
 + i{ p^+ \over 2 \tilde x^+}( \tilde x-\xi)^2 +in\theta \right]
  }
(to derive this, it is convenient to use $\psi_{J} (X) =
e^{\theta J} \, \psi_{J} (\tilde X)$).

The basis functions \vdef\ are singular at $x^+ =0$. More
explicitly, we have
 \eqn\psili{\lim_{x^+ \to 0}V_{p^+,J,n,m^2} (x^+ ,x, x^-,z) =
 \sqrt{{2\pi \ov p^+}} e^{-ip^+x^- + i \left(
 {n-J\over R}\right)z} \, \delta(x-\xi) =
 \sqrt{{2\pi \ov p^+}} e^{-ip^+ \tilde x^- + i n \theta} \,
 \delta(\tilde x-\xi)
 }
This implies that these basis functions are localized at $\tilde
x = x= \xi$ at $x^+=0$ \LMS.

These functions are potentially dangerous to use as vertex
operators for two related reasons:

\item{1.}  Focusing the particles at a point results in infinite
energy density and can cause large back reaction when
the system is coupled to gravity (as in string theory).

\item{2.}  The expression for the basis functions in terms of
plane waves as in the second expression in \invw\ involves
integrating over arbitrarily high energies.  This leads to certain
divergences in S-matrix elements.

The origin of these problems can be traced back to the
diagonalization of $\widehat J$, and one might expect that they
 can therefore be avoided by smearing them with smooth functions of
$J$ of rapid decrease.\foot{Recall that the Schwarz space of
functions of rapid decrease is the space of functions with $
\lim_{J\to \pm \infty} \vert J^n ({d\over dJ})^m f(J)\vert  = 0 $
for any $n,m$. This space is preserved by Fourier transform
\reedsimon.}  We now demonstrate that this is the case. Consider
the wave packets
 \eqn\Udef{U_{p^+,f(J),n,m^2}=\int dJ f(J) V_{p^+,J,n,m^2}}
where $f(J)$ is a  smooth   function of rapid decrease.
Their inner product is
 \eqn\Uinner{\int_{\CF_{x^+}}U^*_{p^+,f,n,m^2}  \, U_{\tilde p^+,
 \tilde f, \tilde n, m^2} = (2 \pi)^3 R \delta (p^+ - \tilde p^+)
 \,\delta_{n,\tilde n} \, \int dJ f(J)^* \tilde f(J) \ .}
Since $V$ is bounded for
$x^+\not=0$, the integral over $J$ in \Udef\ converges for all
$x^+\not=0$.  Moreover the  singularity at $x^+=0$ is smoothed out
 \eqn\Uli{\eqalign{ \lim_{x^+ \to 0}U_{p^+,f,n,m^2} (x^+ ,x,
 x^-,z) =& \sqrt{{2\pi p^+}} \exp\left(-ip^+x^- + i
 (n+xp^+){z\over R} \right) \,f(-x p^+) \cr
 =& \sqrt{{2\pi p^+}} e^{-ip^+ \tilde x^- + i n \theta} \,
 f(-\tilde x p^+)} }
Therefore, $U$ does not have any singularities.  This solves the
first problem mentioned above.

Using \invw\vdef\Udef\ we can express $U$ as an integral over
plane waves
 \eqn\invwU{\eqalign{
 U_{p^+,f,n,m^2} = &\int {dJ dp \over \sqrt{2 \pi p^+} }\ f(J) \cr
 &\exp \left[ip {J \over p^+}-i p^+ x^- - i {p^2 +
 \left({n-J\over R}\right)^2 + m^2 \ov 2 p^+}
 x^+ + i px +i\left({n-J\over R}\right)z  \right]
 }}
Since $f(J)$ is of rapid decrease its Fourier transform is of
rapid decrease.   Therefore, in \invwU\ the support of wave
functions with large energy ${1 \ov 2 p^+} \left(p^2
+\left({n-J\over R}\right)^2 + m^2\right)$ is
suppressed\foot{More precisely, this statement is true only for
fixed $(x^{\pm}, x,z)$, and there is no universal bound for all
values of $(x^{\pm}, x,z)$. }.

It is important that in the parabolic orbifold the label $J$ is
discrete.  Therefore we cannot form wave packets as here and
smear the singularity.  Note that by integrating over $p^+$ the
singularities of the wave functions with $J\not=0$ can be smeared.
However, there is no way to avoid the singularity of the
functions with $J=0$.

To demonstrate this general discussion of wavefunctions in the
null-brane, consider the wave packet with $f(J)= e^{- {J^2 \ov
2}}$,
 \eqn\alphao{
 U_{p^+,- {J^2 \ov 2},n,m^2} = \int dJ \, e^{- {J^2 \ov 2}} \,
 V_{p^+,J,n,m^2}}
It is simple to express the above function in $\tilde x$
coordinates
 \eqn\exfU{\eqalign{
 U_{p^+,- {J^2 \ov 2},n,m^2} =  \sqrt{2 \pi p^+ \ov K} \,
 \exp \left[-{Q \ov 2 K}\right] \, \exp \left(- i p^+ \tilde x^-
 - i {m^2 \ov 2 p^+} \tilde x^+ + i n \theta \right)
 }}
where
\eqn\defpar{\eqalign{
 K & =  p^+ \tilde x^+  + i \left[\left({\tilde x^+\over R}
 \right)^2 - 1 \right] , \cr
 Q & = \left({p^+\tilde x^+ \over R^2}- i  (p^+)^2 \right) \tilde x^2
 + \left[{\tilde x^+ \ov p^+ R^2} - i  \left({\tilde x^+\over R}
 \right)^2 \right] n^2
   + {2 \tilde x^+ \tilde x n\over R^2} \ . \cr
 }}
It is manifest in \exfU\ and \defpar\  that $U$ is completely
regular everywhere. It is also instructive to write $U$ as a wave
packet in momentum space using the second line of \invw. We find
that \eqn\exfpU{\eqalign{
 U_{p^+,- {J^2 \ov 2},n,m^2} =  \sqrt{1 \ov p^+
 +i {\tilde x^+ \ov R^2}} \,
 \exp \left(- i p^+ \tilde x^- - i {m^2 + {n^2 \ov R^2}
 \ov 2 p^+} \tilde x^+ + i n \theta \right)  \,
 \int d \tilde p  \, g (\tilde p) \, e^{i \tilde p
 \tilde x - i {\tilde p^2 \ov 2p^+} \tilde x^+}
 }}
with
 \eqn\defmpa{
 g(\tilde p) = \exp \left[ - {1 \ov 2 \left(1
 + i {\tilde x^+ \ov p^+ R^2}\right)}
 \left({\tilde p \ov p^+}  + {\tilde x^+ n \ov p^+ R^2}
 \right)^2\right] \ .
 }
Thus $U$ corresponds to a Gaussian wave packet in $\tilde p$
which is the momentum conjugate to $\tilde x$. In particular, the
higher energy region of the integration is suppressed.

The $R\to 0$ limit of the null-brane is the parabolic orbifold
times a noncompact line parametrized by $z$.  In order for the
function $V_{p^+,J,n,m^2}$ to have a sensible limit we should
hold the momentum along the $z$ direction, $k={n-J\over R}$,
fixed as $R\to 0$.  The resulting function is
$\psi_{p^+,J,k,m^2}$ of \invw\ with $J=n\in \IZ$, which is a good
function on the parabolic orbifold times the line. In order to
have a good limit for the wave packets $U$ \Udef\ we should let
the function $f(J)$ depend also on $n$ and $R$, $f_n(J,R)$, such
that $\lim_{R\to 0} f_n(J=n-kR,R)=\tilde f_n(k)/R$ with finite
$\tilde f_n(k)$.  Then
 \eqn\ulim{\lim_{R\to 0}U = \int dk \tilde f(k)
 \psi_{p^+,J=n,k,m^2}}
i.e.\ it has fixed value for $J=n\in \IZ$ and is a wave packet in
$z$.  Clearly, this function diverges at $x^+=0$.

One can work out the propagator for a massive particle on the
null-brane. It can be obtained from  the propagators in
$\IR^{1,3}$  by summing over the images under the orbifold
action, leading to
 \eqn\propo{ G_F (\CP_1, \CP_2) = {i \ov 8 \pi} \sum_{n\in \IZ}
 \biggl({m^2 \over \Delta_n + i \ep } \biggr)^{1/2}
 H^{(2)}_1(\sqrt{m^2 (\Delta_n + i \ep) }) }
where $\Delta_n$ is the invariant distance square between a point
$\CP_1$ and the $n^{th}$ image of $\CP_2$ \newdetn.  Note that the
propagator for the parabolic orbifold is also given by \propo\
with $R=0$ in \newdetn.

The propagator \propo\ can  be expanded in terms of the wave
functions discussed in this section,
 \eqn\propag{\eqalign{
 G (\CP_1, \CP_2)  = \theta (\Delta y^+) \, D (\CP_1, \CP_2)
 + \theta (- \Delta y^+) \, D (\CP_2, \CP_1) }}
where when $\Delta y^+ > 0$
 \eqn\defdpro{ \eqalign{
 & D(\CP_1, \CP_2)  = \int_0^\infty
 {dp^+ \over (2\pi)^3 2p^+ R} \int_{-\infty}^{\infty} dJ \,
 \sum_{n\in \IZ} {1 \over \sqrt{i y_1^+} \sqrt{- i y_2^+}} \cr
 & \times \,
 \exp \left[-ip^+ \Delta y^- + i J \Delta y + i {n -J \ov R}
 \Delta z  -i{J^2\over 2p^+} \left(-{1\over y^+_1} + {1 \over
 y^+_2}\right) - i {m^2 + \left({n - J \ov R}\right)^2
 \over 2p^+} \Delta y^+  \right] \cr
 }}
Note that the sum over $n$ in \defdpro\ and \propo\ are not the
same. They are related by a Poisson resumation.

\newsec{String quantization on the null-brane}

In this section we consider the canonical quantization of strings
in the null-brane background.  The discussion is very similar to
that of the parabolic orbifold in \LMS\ and therefore it will be
very brief.

In the covariant formalism we use four free fields $x^\pm, x,
z$.  In the $w$ twisted sector they are subject to the twisted
boundary conditions
 \eqn\boundarycon{\pmatrix{x^+(\sigma+2\pi,\tau)\cr
 x(\sigma+2\pi,\tau)\cr
 x^-(\sigma+2\pi,\tau)\cr
 z(\sigma+2\pi,\tau)\cr} = g_0^w \cdot \pmatrix{X\cr z\cr} =
\pmatrix{x^+(\sigma,\tau)\cr
 x(\sigma,\tau)+ 2\pi w x^+(\sigma,\tau)\cr
 x^-(\sigma,\tau)+ 2\pi w x(\sigma,\tau) + \half (2\pi w)^2
 x^+(\sigma,\tau)\cr
 z(\sigma,\tau) + 2\pi w R\cr}}
These can be ``solved'' using free fields \LMS\ and lead to an
interesting exchange algebra similar to exchange algebras in RCFT%
\foot{Although there is a potential role for noncommutative
geometry here, note that the spacetime is perfectly Hausdorff, as
is the WZW theory.}.

In the lightcone gauge we have two possible procedures:
\item{1.} Use the original $x^\pm,x,z$ coordinates.  The advantage
of this method is that the worldsheet Lagrangian is free.  The
somewhat unusual aspect of this procedure is that the worldsheet
Hamiltonian $p_x^-$ is not invariant under the orbifold action
which is a gauge symmetry of the system.  Also, in the twisted
sectors the periodicity rules of the fields depend on the
worldsheet time $x^+=\tau$.
\item{2.} Use an invariant Hamiltonian.  This can be done either
with the coordinates \newcoordii\zud, where the Hamiltonian is
$p_{y}^-=i\partial_{y^+}$ or with the coordinates \gloco, where
the Hamiltonian is $p_{\tilde x}^-=i\partial_{\tilde x^+}$.  The
disadvantage of this procedure is that unlike the first one, the
system is time ($x^+$) dependent.

It is essential to keep as dynamical variables in the Lagrangian
the zero modes of $x^-$ in the first procedure, and the zero mode
of $y^-$ or $\tilde x^-$ in the second procedure
\refs{\LMS,\longpaper}. In the first procedure this coordinate is
needed to ensure the invariance of the system under the orbifold
action.  In the second procedure this is needed in order to
absorb all infinite renormalization constants in appropriate
counter terms.

For brevity here we will follow only the first procedure.  The
light-cone gauge Lagrangian is
 \eqn\lcgii{
 L= - p^+ \partial_\tau x^-_0 + {1\over 4\pi \alpha'}
 \int_0^{2\pi}\!\! d\sigma \left(\alpha' p^+ \p_\tau x \p_\tau x
 - {1\over \alpha'p^+ } \p_\sigma x \p_\sigma x
 + \alpha' p^+ \p_\tau z \p_\tau z
 - {1\over \alpha'p^+ } \p_\sigma z \p_\sigma z\right)
 }
Invariance under constant shifts of $\sigma$ is implemented by
imposing \longpaper
 \eqn\lzmlzb{\int d\sigma \left(\partial_\sigma x \partial_\tau
 x + \partial_\sigma z \partial_\tau
 z- {1\over 2\tau} \partial_\sigma x^2 \right)=0}
It is important that the Lagrangian \lcgii\ and the expressions
for the constraint \lzmlzb\ are invariant under the orbifold
identification
 \eqn\identlcg{\eqalign{
 &x(\sigma,\tau) \to x(\sigma,\tau) +2\pi n \tau\cr
 &x^-_0 (\tau) \to x^-_0(\tau) + 2\pi n \int_0^{2\pi}
 {d \sigma \over 2\pi}
 x(\sigma,\tau) +{(2\pi n)^2\over 2} \tau \cr
 &z(\sigma,\tau)\to z(\sigma,\tau) + 2\pi nR}}
The equation of motion for $x^-_0$ sets $p^+$ to a constant. The
equation of motion for $p^+$ leads to
 \eqn\lcgii{\eqalign{
 &P_{x^-}=p^+\partial_\tau x^-_0 = {1\over 4\pi\alpha'}
 \int_0^{\ell} d\sigma \left( \p_\tau x \p_\tau x
 +\p_\sigma x \p_\sigma x  + \p_\tau z \p_\tau z
 +\p_\sigma z \p_\sigma z \right)\cr}}
where we have rescaled $\sigma$ to range in $[0,\ell=2\pi
\alpha'p^+)$.

A complete set of solutions to the equations of motion in the
$w$-twisted sector, can be expressed in terms of harmonic
oscillators:
 \eqn\xysol{\eqalign{ &
 x(\sigma,\tau) =-{J\over p^+} + {p \over p^+}\tau + {2\pi w
  \sigma \tau\over \ell}+ \cr
 & \qquad i \left({\alpha'\over 2}\right)^\half \sum_{n\not= 0}
 \left\{ {\alpha_n \over n} \exp \left[-
 {2\pi in(\sigma+\tau) \over \ell}\right] + {\tilde \alpha_n \over
 n} \exp\left[ {2\pi in(\sigma-\tau) \over \ell}\right]
 \right\}\cr
 & z(\sigma,\tau) =z_0 + {k \over p^+}\tau + {2\pi w R
  \sigma \over \ell}+ \cr
 & \qquad i \left({\alpha'\over 2}\right)^\half \sum_{n\not= 0}
 \left\{ {\alpha_n^z \over n} \exp \left[-
 {2\pi in(\sigma+\tau) \over \ell}\right] + {\tilde \alpha_n^z
 \over n} \exp\left[ {2\pi in(\sigma-\tau) \over \ell}\right]
 \right\}\cr}}
The solution of $x_0^-$ is obtained from \lcgii.  Upon
quantization these oscillators obey the standard canonical
commutation relations and $n=J+kR,w\in \IZ$.

As in \LMS, it is straightforward to extend the worldsheet
Lagrangian \lcgii\ to the Green-Schwarz formalism.  For
concreteness consider the model on $\CO\times \IR^6$.  Before
taking the quotient by $\Gamma$ we should add to \lcgii\ six free
worldsheet bosons $x^i$ and eight rightmoving fermions $S^a$ (in
the type II theory we also need eight leftmoving fermions and in
the heterotic string also leftmoving degrees of freedom for the
internal degrees of freedom).  It is easy to see using the
symmetries of the problem that after the action by $\Gamma$ the
added fields $x^i$ and $S^a$ remain free.  The boundary
conditions of these fields depend on the spin structure around
the nontrivial cycle.  For the spin structure which preserves
supersymmetry $x^i$ and $S^a$ are periodic around the string. For
the other spin structure $S^a$ transforms with $(-1)^w$, as in
the supersymmetry breaking compactification of \rohm.

\newsec{Torus partition function}

In the one-loop amplitudes we sum over the ``sectors''
\eqn\oneloop{ \eqalign{ (X,z)(\sigma^1 + 1, \sigma^2) & = \left(
e^{2 \pi  w^a \CJ} X, z + 2 \pi R w^a\right)(\sigma^1, \sigma^2)
\cr (X,z)(\sigma^1 , \sigma^2+1) & = \left(e^{ 2 \pi  w^b \CJ} X,
z + 2 \pi R w^b \right)(\sigma^1, \sigma^2) \cr} }
Comparing with the analogous calculation in \LMS, the only
difference is the replacement $(x^+)^2 \to (x^+)^2 + R^2$ as the
effective radius-square in equation (5.13) of \LMS. That is, the
contribution of the classical action to the torus amplitude is
given by
\eqn\replacess{ \exp\Biggl[-  {\pi [(x^+)^2 + R^2]\over \ap }
{\vert w_b + w_a \tau\vert^2 \over \tau_2} \Biggr] }
Equation \replacess\ is most easily derived using the coordinate
system $(\tilde X, \theta)$ since the classical configuration is
then given by
 \eqn\newsolsec{ \theta(\sigma^1, \sigma^2) =2\pi (w^a
 \sigma^1 + w^b \sigma^2) }
with $\tilde X$ a constant.

We would like to make a few comments about the torus amplitude:
\item{1.}   Both in the parabolic orbifold and in the null-brane we
can study the system either with supersymmetric boundary
conditions around the nontrivial cycle or with nonsupersymmetric
boundary conditions as in \rohm.  With supersymmetric boundary
conditions we find cancellations and the one loop cosmological
constant vanishes.
\item{2.}  The final answer is a cosmological constant as a
function of $(x^+)^2+R^2$.  Regardless of the boundary conditions
as $R\to \infty$ we recover the standard flat space answer
because in this limit our spacetime becomes time independent.
Similarly, in the large lightcone time limit $x^+\to \pm \infty$
the cosmological constant vanishes in the superstring theory.
\item{3.}  As $R\to 0$ we recover the results of \LMS\ for the
parabolic orbifold.
\item{4.}  If $R$ is sufficiently small, $(x^+)^2+R^2$ can be
smaller than order $\alpha'$.  Then if supersymmetry is broken
the system can have tachyons in the winding sector and the
cosmological constant can diverge.
\item{5.}  As $x^+\to 0$ for finite $R$ we do not find a continuum
of winding modes because the T-dual space is of finite size
($\alpha'/R$).

\newsec{Tree amplitudes}

In this section we consider the tree-level amplitudes for the
untwisted modes in the null-brane. We will show that  the
tree-level amplitudes for the wave packets $U$ of \Udef\ in the
null-brane are better behaved than the amplitudes studied in
\LMS. The amplitudes involving twisted modes, which are important
for understanding twisted mode production, will be left for
future work.

As in \LMS, we  calculate the tree-level untwisted amplitudes in
the null-brane using the inheritance principle from those in flat
space in the $J$-basis by restricting $n=J + kR$ to integers.
Therefore, in the $J$-basis the tree level untwisted S-matrix is
essentially the same in flat space, in the parabolic orbifold and
in the null-brane. The amplitudes in the $J$-basis in flat
spacetime are in turn computed from those in the momentum basis by
a Fourier transform, thanks to the simple relation between the
$J$-basis functions and the plane waves.  For simplicity we will
only look at tachyon amplitudes.

The vertex operator for the tachyon is given by
 \eqn\verJ{ V_{p^+,J,n,p_{\perp}} (\sigma,\tau) =
 {1 \over \sqrt{2 \pi
 p^+}}\ \int_{-\infty}^\infty dp \,
 e^{ip \xi} \, e^{i \vec{p} \cdot \vec{X}(\sigma,\tau)},
 \qquad \xi=-{J \ov p^+}}
where
 \eqn\veropa{e^{i \vec{p} \cdot \vec{X}(\sigma, \tau)}
 = \exp \left[-i p^+ x^- - i p^- x^+ + i p x + i k z+
 i \vec{p}_{\perp } \cdot \vec{x}_{\perp} (\sigma, \tau) \right]}
($\vec p_{\perp }$ and $\vec x_{\perp }$ denote vectors in other
transverse dimensions) is the standard on-shell tachyon vertex
operator with
 \eqn\momdefa{\eqalign{& p^- = {p^2  + \m^2 \ov 2
 p^+}, \qquad \m^2= m^2  + k^2 + \vec{p}_{\perp }^2,\qquad \cr
  & m^2 = - {4 \ov \apr}, \qquad k = {n - J \ov R} \ .
 }}

It then follows that in the null-brane the three- and four-point
amplitudes  in terms of the above vertex operators are given
precisely by equations (6.4) -- (6.6) and (6.13) -- (6.17) of
\LMS\ with the substitution
 \eqn\subst{\eqalign{
 (\vec p_{\perp i})_{\rm there} = (k_i, \vec p_{\perp i}), \qquad
 k_i = {n_i - J_i \ov R} \ . }}
Thus the behavior of the four-point amplitudes are the same as
those discussed in \LMS, which involve integrating the momentum
space amplitudes over an infinite  range of the  Mandelstam
variable $s$. An important question is whether the S-matrix is
finite, since the integrals involve very large $s$, i.e.
potentially infinite center of mass energy. Roughly, the
contribution of  the large $s$ region of the integral to the
amplitudes can be written as $$A_{J} \approx I_+ + I_- $$ with
 \eqn\vadel{\eqalign{
   \qquad I_{\pm}  \sim \int_{-\infty}^{\infty}  {d \sig \ov
   |\sig|} \,
  e^{i \sig F_{\pm} (J_i,p_i^+)} \; A_{VS} (s(\sig),t_{\pm}
  (\sig),u_{\pm} (\sig))
  }}
where $A_{VS}$ is the Virasoro-Shapiro amplitude in momentum
space, $F_{\pm} (J_i,p_i^+)$ are functions of $J$ and $p^+$ of the
external particles, and
 \eqn\exprfors{
 s = (p_1^+ + p_2^+)  \sig^2 +
 \left(\sqrt{p_1^+ \ov p_2^+} k_1 -  \sqrt{p_2^+ \ov p_1^+}
 k_2\right)^2
 + \mu_s (\vec p_{\perp i},p_i^+) \ .}
$\mu_s$ is a function of $\vec p_{\perp}$ and $p_i^+$ only, whose
detailed form will not concern us below. Note that in \vadel\ the
Mandelstam variables $t,u$ are different functions of $\sig$ in
$I_{\pm}$, given by $t_{\pm} (\sig)$ and $u_{\pm} (\sig)$
respectively. For more detailed expressions of various functions
in \vadel\  see \LMS. To compare with the expressions in \LMS,
note that $\sig$ in \vadel\ corresponds to $q_{+}$ in equation
(6.17) there and $I_{\pm}$ here correspond to equation (6.16) of
\LMS\ in the $|q| \to \infty$ and $|q| \to 0$ limits respectively.

The convergence behavior of \vadel\ can be summarized as follows:

\item{1.}  For generic $p_i^+$, $A_{VS} (s(\sig),t_{\pm}
(\sig),u_{\pm}(\sig))$ behaves for large $s$ as in the hard
scattering limit of large $s,t,u$ with fixed ratios.  In this
limit
 \eqn\larges{A_{VS} \sim e^{-\lam s} }
for some positive constant $\lam$. The integral over $\sig$
\vadel\ converges for all values of $J_i$ and in particular the
dependence on $J_i$ is analytic.

\item{2.} When $p_t^+ =p_3^+ - p_1^+ =0$, while $I_-$ behaves as
in $1.$ above and is finite, $I_+$ is divergent for some values of
$J_i$ and $\vec p_{\perp i}$. More explicitly, as $|\sig|$ goes to
$\infty$,
 \eqn\finnt{
 t_+ \approx -(k_1 - k_3)^2 - (\vec p_{\perp 1} - \vec
 p_{\perp 3})^2 = -(k_1 - k_3)^2 - \vec p_{\perp t}^2}
remains finite. Thus the integrand of $I_+$ in \vadel\ is in the
Regge scattering limit of large $s$ with finite $t$. In this
limit, the leading terms in $I_+$  can be  written as
 \eqn\ampreg{I_+ \sim  \int  {d \sig \ov |\sig|}  \,
 e^{i \sig F_+} \; s^{-\ha \apr \m_t^2}
 \, \left(\Ga \left({\apr \ov 4} \m_t^2 \right) \right)^2 \,
 \sin \left({\apr \pi \ov 4} \m_t^2 \right) \,}
where $s$ is given by \exprfors\ and
 \eqn\mtdef{\apr \m_t^2 = - 4 - \apr t_+, \qquad F_+ = {1 \ov
 \sqrt{\mu_{12}}}( J_3 -
 J_1), \qquad \mu_{12} = {p_1^+ p_2^+ \ov p_1^+ + p_2^+}
   \ .}
It then follows that the integral for $A_J$ diverges as $|\sig|
\to \infty$ when
 \eqn\conddi{p_t^+=0; \qquad J_t=J_3 -J_1 =0 ; \qquad
 \apr\vec p_{\perp t}^2 + \apr(k_1 - k_3)^2 < 4 }
i.e.\ close to forward scattering (but not only in the forward
direction).  There is a similar divergence with $3\leftrightarrow
4$ associated with the u-channel.

In the null-brane, as we discussed in section 3, instead of using
singular functions \verJ\ as vertex operators, it is more
appropriate to form wave packets \Udef\ in terms of rapidly
decreasing functions of $J$. We showed that the wave packets are
finite everywhere and in particular, the high energy region of the
integration is suppressed by a rapidly decreasing function. It is
clear that when $p_t^+ = p_3^+ - p_1^+ \neq 0$, by integrating the
amplitudes \vadel\ over $J_i$ with a kernel $\prod_i f_i(J_i)$ we
find a finite answer. The same is true for $I_-$ when $p_t^+=0$.
For $I_+$, when $p_t^+=0$, the situation is more subtle. First
note that since
 $$
 k_i = {n_i - J_i \ov R}, \quad n_i \in {\bf Z}
 $$
when we integrate over $J_i$ from $-\infty$ to $\infty$,
effectively we are integrating $k_i$ over the same ranges. At
first sight this seems to invalidate our approximation used in
\ampreg\ ($t_+$ is no longer fixed). However, since $f_i(J_i)$
suppresses large $J_i$ (i.e. large $k_i$) contribution, we expect
that for the purpose of estimating the convergence of the
amplitudes, it is still legitimate to use \ampreg. Therefore we
have the amplitudes
 \eqn\inte{
A_{U} \sim  \prod_{i=1}^4 \left( \int d J_i \, f_i(J_i) \right) \,
I_+ + \cdots
 }
where $I_+$ is given by \ampreg\ and $\cdots$ denotes other finite
contributions including those from $I_-$. The integrals \inte\ are
potentially divergent around $J_t = J_3 - J_1 \approx 0$ when $n_1
= n_3$ and $\vec p_{\perp t} \approx 0$.  In this region, using
\ampreg\ and noting that $\alpha' t_+ \cong 0$,
\inte\ can be approximated as
 \eqn\finapp{
 A_U \sim \int d J_t \! \int  {d \sig \ov |\sig|} \;
 {1 \ov {J_t^2 \ov R^2}  +  \vec p_{\perp
 t}^2} \, e^{i {\sig \ov \sqrt{\mu_{12}}}
 J_t} \,
 |\sig|^{4 - {\apr \ov R^2} J_t^2 - \apr \vec p_{\perp
 t}^2} \
 }
for a reasonably general class of rapidly decreasing functions
$f_i$.\foot{Note that \finapp\ is valid even when the support of
$f_1(J_1)$ is very different from that of $f_3(J_3)$, as long as
$f_1 f_3$ does not vanish for $J_1 = J_3$.} The factor ${1 \ov
{J_t^2 \ov R^2}  +  \vec p_{\perp t}^2}$ in \finapp, which comes
{}from the product of the $\Gamma$ and the sine functions in
\ampreg, can be understood as arising from a $t$-channel
propagator of a soft graviton. Interestingly, when $\vec p_{\perp
t} =0$, the integrand of the $J_t$ integral is singular at
$J_t=0$ due to both an IR divergence coming from the $1/J_t^2$
factor and a UV divergence coming from the unbounded $\sig$
integral.

If we define the integral \finapp\ by first integrating over
$J_t$ and then over $\sigma$ we find that it is finite whenever
$\vec p_{\perp t} \neq 0$. This can be understood from the fact
that the potential UV divergence from the $\sig$ integral is
integrable when we integrate over $J_t$.  However, the familiar
$1/\vec p_{\perp t}^2$ singularity at $\vec p_{\perp t}=0$ is
enhanced by the UV region of large $\sigma$ to a stronger
singularity -- a larger inverse power of $ \vec p_{\perp t}^2$ up
to logarithmic corrections.

To illustrate this more clearly, let us look at the $\apr \to 0$
limit of \finapp. In this case the integrals can be easily
evaluated, and we find that the amplitude scale as $ {1 \ov |\vec
p_{\perp t}|^5} $. $\apr$ corrections make the amplitudes less
divergent but not finite. A more precise analysis of the
singularities would be desirable.

To summarize, we find that in the null-brane, the $J$-basis
amplitudes \vadel\ suffer from the same divergences as those in
the parabolic orbifold. When we use smooth wave packets \Udef\ in
terms of rapidly decreasing functions of $J$, the amplitudes are
better behaved,  but still have divergences when $p_t^+ =n_3
-n_1=\vec p_{\perp t} =0$.

\bigskip
\medskip

\centerline{\it Discussion of various divergences}

\medskip

There are many known  examples of divergent S-matrix elements at
non-generic kinematics.  Some standard examples include
singularities associated with on shell intermediate particles.
In terms of the integral over the first quantized parameters
(moduli) they originate from the regions $x^+ \to \pm \infty$, in
which intermediate propagators are long.  In our case, since our
external states are not $p^-$ invariant (corresponding to the
lack of $p^-$ invariance of the parabolic orbifold and the
null-brane backgrounds), the corresponding singularities in the
S-matrix are not poles.  Instead, our S-matrix exhibits some
other nonanalytic dependence on the external momenta which
originates from the large $x^+$ behavior (IR singularities) \LMS.

The singularities of the four-point amplitudes in the $J$-basis
when $p_t^+=J_t=0$ and $\vec p_{\perp t}^2$ is sufficiently small
( as in \conddi) are IR singularities because $p_t^+=0$, but they
also have another crucial element. They arise from a divergence of
\ampreg\ at large $s$, and therefore are also UV singularities.
They originate from the singularity of $V_{p^+,J,n,\vec
p_{\perp}}$ at $x^+ =0$, or equivalently from the fact that
$V_{p^+,J,n,\vec p_{\perp}}$ has arbitrarily large energy.   As
discussed after equation \psili, the focusing at $x^+ =0$ leads
to infinite energy density and therefore large coupling to the
graviton, and hence the divergence.

Unlike the standard IR singularities, our singularities occur for
a {\it range } of values of $\vec p_{\perp t}^2$ and not only for
$\vec p_{\perp t}=0$.  Therefore, it is impossible to remove them
by the standard procedure of dealing with IR singularities, and
the results suffer from incurable IR divergences.  Presumably this
signals breakdown of the perturbation theory\foot{It would be
rather desirable to have a clean computation of 1-loop scattering
amplitudes indicating a clear divergence.}. This breakdown of
perturbation theory should perhaps be interpreted physically as
follows: If we try to scatter particles which become infinitely
focused, the nonlinear backreaction of the metric results in a
singularity. The condition $p_t^+=J_t=0$ states that the incoming
particle $1$ and the outgoing particle $3$ are focused at the
same point at $x^+=0$. It is clear that otherwise, the amplitude
cannot diverge \LMS.  For a related recent discussion see
\refs{\lawrence,\hp}.

In order to solve this problem we need to prevent the focusing of
the incoming particles.  This can be done only by scattering wave
packets which smear the values of $J_1$ and $J_2$.

In flat space this is easily done.  For example we can scatter $p$
eigenstates. More generally, a wave packet in the $J$-basis is
equivalent to a wave packet in momentum space by a Fourier
transform and therefore the amplitudes are finite.

In the parabolic orbifold since $J$ is discrete we do not have the
freedom to integrate over $J$ to form wave packets.  This
reflects the fact that in the parabolic orbifold, the energy of
incoming states is  always blue-shifted without bound as they
approach the singularity.  Therefore, the S-matrix elements in
the parabolic orbifold cannot be computed in perturbation
theory.  It might also mean that the notion of an S-matrix for
scattering from $x^+=-\infty$ to $x^+=+\infty$ is simply not
valid in the interacting theory due to large back reaction
effects.  It is possible that one should formulate the theory
with only one of the two asymptotic regions and interpret
vertex operator correlators differently.

Another perspective on the divergence comes from noting that the
divergence arises from Regge behavior of the scattering
amplitudes. Already in the standard flat space amplitudes  Regge
behavior is related to a breakdown of string perturbation theory
for forward scattering amplitudes. There are plausible
resummation techniques for determining the nonperturbative
forward amplitudes \refs{\soldate,\amati}. If we (naively!) apply
the inheritance principle to such resummed amplitudes we find that
the divergence is softened.

In the null-brane, the spacetime is smooth everywhere and we
expect that, as in flat space, the amplitudes for wave packets
should be well-behaved. Indeed,  we find,  by integrating the
amplitudes \ampreg\ over $J_i$ with a kernel $\prod_i f_i(J_i)$, a
finite answer even when $p^+_t=0$ for generic $\vec p_{\perp i}$.
However, when $p_t^+=n_3 - n_1=0$ and $\vec p_{\perp t} \to 0 $
the amplitudes diverge like an inverse power of $\vec p_{\perp
t}^2$. As we have emphasized, this singularity is not merely the
standard IR singularity due to the propagator of the soft graviton
in the {\it t-channel}.  It is enhanced due to the integration
over the UV region of the {\it s-channel}. It would be desirable
to have a clearer understanding of this curious mixing of UV/IR
divergences. Unlike the situation in the parabolic orbifold or in
the scattering of $J$ eigenstates in the null-brane, which we
discussed above, here the singularity occurs only in the forward
direction, i.e.\ for $p_t^+=\vec p_{\perp t} = 0 $.  We will not
try to prove it here, but we suspect that, when there are a
certain number of uncompact transverse dimensions, such a
singularity may be removed by the standard procedure for dealing
with IR divergences due to soft gravitons, and therefore the
physical results are sensible.

There is an important special case of the previous discussion in
which there are no transverse noncompact dimensions to the null
brane.  Then $\vec p_\perp$ takes discrete values, and therefore
the divergence at $\vec p_{\perp t}$ cannot be removed.  We
expect that in this case the singularity is just as harmful as in
the parabolic orbifold, and that perturbation theory breaks down.
It might come as a surprise that in the null-brane, when there
are no other noncompact dimensions, even with a smooth geometry
and smooth wave packets, we find divergences in the amplitudes
which appear to be incurable. This situation is reminiscent of
the singular response of de Sitter space to small perturbations
in the infinite past (see e.g. \refs{\BanksFE,\WittenKN}).

Let us return to the null-brane with sufficient number of
noncompact dimensions. It is of some interest to understand how
the finite $S$-matrix at $R>0$ becomes singular for $R\to 0$. As
$R\to 0$ the wave packets $U$ have fixed $J=n\in \IZ$ and become
wave packets in $z$ \ulim.  The limit of the S-matrix has the
same divergence we encountered in the parabolic orbifold.  As in
the parabolic orbifold, this can be understood either as a result
of the singularity of the vertex operators at $x^+=0$, or as a
consequence of the fact that they have arbitrarily large energy.

In conclusion, we found interesting singularities in S-matrix
elements, which are ``UV enhanced IR divergences.'' We believe
that their most likely interpretation is associated with large
back reaction of the geometry.  In the parabolic orbifold (and in
the null-brane without sufficient number of noncompact
dimensions) they signal breakdown of perturbation theory.  A
similar problem exists in the null-brane (and in flat space) for
vertex operators with fixed $J$. However, the S-matrix of the
wave packets $U$ of \Udef\ in the null-brane with some noncompact
dimensions appears to be consistent. Thus these models provide
good laboratories for studying aspects of string perturbation
theory in a time-dependent setting  and should be investigated
more thoroughly.

\newsec{Some general comments on the class of free-field
time-dependent orbifolds}

\subsec{Criteria for nonsingular physics}

In this section we make some preliminary remarks on the
classification of those time-dependent string orbifold models to
which the the methods of \LMS\ can be straightforwardly applied.
We consider spacetimes of the form $(\IR^{1,d+1}/\Gamma) \times
\CC^\perp$ where $\Gamma\subset \CP(1,d+1)$ is a discrete subgroup
of the Poincar\'e group $\CP(1,d+1)$ in $\IR^{1,d+1}$ and
$\CC^\perp$ stands for some transverse spacetime (more precisely,
some transverse conformal field theory) on which $\Gamma$ does
not act.

In order to state any classification result one must formulate
carefully the list of desired properties. We would like to have:

\item{1.} Time-dependence
\item{2.} Time-orientability
\item{3.} Some unbroken supersymmetry
\item{4.} No CTC's

Unbroken supersymmetry and time-dependence implies that there is
a null Killing vector  and hence spacetimes satisfying these
criteria are foliated by hypersurfaces $\CF_{x^+}$ of constant
$x^+$, where $\p_{x^-}$ is null \figueroai. A further criterion
one might wish to impose is:

\item{5.} For generic $x^+$ there are no closed null curves
(CNC's).

Somewhat surprisingly,  these criteria (and even the stronger
condition that for {\it all} $x^+$ there are no closed causal
curves) do not guarantee nonsingularity of the physics. Things
can go wrong if there is a family of homotopically inequivalent
spacelike curves $\gamma_n$ whose lengths $L(\gamma_n)$ approach
zero for $n\to \infty$. We refer to such a family as {\it nearly
closed null curves} (NCNC's). To eliminate this pathology   we
might also wish to impose:

\item{6.} There is a positive constant $\kappa$ such that all
closed homotopically nontrivial curves have length
$L(\gamma)>\kappa$.

\subsec{Classification result for $d=2$}

In this case one can identify spacetime with $2\times 2$
Hermitian matrices
\eqn\sti{ \bx = \pmatrix{\sqrt{2} x^- & x \cr \bar x  & \sqrt{2}
x^+\cr} }
where $x\in C$, $x^\pm \in \IR$. The Minkowski metric is
\eqn\minmet{ ds^2  = - \det (d \bx) } and the action of the
Poincar\'e group is
\eqn\poinact{ \bx \to g \bx g^\dagger + \ba }
where $g\in SL(2,\IC)$. We denote elements of the Poincar\'e group
by $\Lambda = (g,\ba)$.

We now classify the possible orbifold groups $\Gamma$. It is not
difficult to show that any discrete group satisfying the criteria
1 -- 4 above can be conjugated into a subgroup of the continuous
group $\CS\subset \CP(1,3)$ defined by elements of the form
\eqn\sgroup{ \Lambda = \biggl( g=  \pmatrix{1&  \xi/\sqrt{2} \cr
0 & 1\cr} ; \ba = \pmatrix{\sqrt{2} a^- & a \cr \bar a & 0 \cr}
\biggr) \qquad \xi,a\in \IC, a^-\in \IR }
Every element of $\CS$ can be written as $t(a^-)\Lambda(\xi,a)$
with
 \eqn\lambdaxi{ \Lambda(\xi, a) := \biggl( g=  \pmatrix{1&
 \xi/\sqrt{2} \cr 0 & 1\cr} ; \ba = \pmatrix{ 0 & a \cr \bar a & 0
 \cr} \biggr) \qquad \xi, a\in\IC }
and
 \eqn\tzee{ t(a^-):= \biggl( g= 1 ; \ba = \pmatrix{ \sqrt{2}
 a^- & 0 \cr 0  & 0 \cr} \biggr) \qquad a^-\in \IR }
{}From the identities:
 \eqn\identone{ t(a^-) \Lambda(\xi,a) = \Lambda(\xi, a) t(a^-) }
 \eqn\identtwo{ \Lambda(\xi_1,a_1) \Lambda(\xi_2,a_2) =
 t(\re(\xi_1\bar a_2)) \Lambda(\xi_1+\xi_2, a_1+a_2) }
it follows that,  group-theoretically, $\CS$ is the
five-dimensional Heisenberg group.

If we also impose criterion 5 above then $\Gamma$ must not contain
any group elements of the form $t(a^-)$ and moreover must  be an
Abelian subgroup of $\CS$. This will be true if $\re(\xi_1 \bar
a_2 - \xi_2 \bar a_1)=0$ for all pairs of elements.

We may now organize the classification by the minimal number $r$
of generators of $\Gamma$.

If $r=1$, then, using conjugation by elements of $\CP(1,3)$ it is
easy to show that the most general possibility is the null-brane,
together with the parabolic orbifold.\foot{This case is very
closely related to the discussion  of Figueroa-O'Farrill and
Sim\'on \fluxbranes. In our language, they classified 1-generator
models for $\IR^{1,10}$. Although their assumptions were more
restrictive than ours, the outcome is the same, at least, for
$\IR^{1,3}$. } That is, we can take $g_1 = \Lambda(2\pi, i R)$
for $R\in \IR$. When $r\geq 2$ we take orbifolds of the null
brane. When $r=2$ the inequivalent possibilities for the second
generator $g_2 = t(a^-_2)\Lambda(\xi_2,a_2)$  fall into two cases
depending on whether or not we can conjugate $a^-_2$ to
zero\foot{At this point we have not yet imposed the condition
that $\Gamma$ is Abelian.}:
\item{1.} $(a^-_2\in \IR^*, \xi_2 \in \IR , 2\pi a_2 = i R\xi_2 )$
\item{2.} $(a^-_2=0, \xi_2 ,a_2 \in \IC  ) $.

The two-generator  model with generators $\Lambda(\xi_i, a_i)$,
$i=1,2$, has an interesting geometry exhibiting the NCNC's
mentioned above.  The action of the group on $x$ is given by
$x\to x+ n_i \omega_i$ with $\omega_i = \xi_i x^+ + a_i$. Thus
the leaves of the foliation $\CF_{x^+}$ are themselves foliated by
tori with time-dependent periods $\omega_i$. If $\omega_1,
\omega_2$ are not rationally related there are no CNC's and no
fixed points. Nevertheless, the physics is potentially singular.
The reason is that at (generically two) critical  times $x^+_{c}$
the tori degenerate, and $\omega_2(x^+) = \lambda \omega_1(x^+)$
for a real number $\lambda$.  When $\lambda$ is rational there
are CNC's. If $\lambda$ is irrational then by Dirichlet's theorem
there is an infinite set of distinct solutions in relatively prime
integers $(p,q)$  to $\vert p\lambda - q\vert < 1/\vert q\vert$
and the geodesics that lift to $q\omega_1 - p\omega_2$ on the
covering space form   an infinite set of closed spacelike
geodesics, $\gamma_n$, in nontrivial homotopy classes, whose
lengths $L(\gamma_n)$ approach zero for $n \to \infty$. Then, for
example, the classical soliton sum (with $\ap=1$)
 \eqn\oneloopsum{
 \sum_{n_i^a, n_i^b\in Z} \exp\Biggl\{ - {\pi \over \im \tau} \vert
 (n_1^a - \tau n_1^b)(\xi_1 x^+ + a_1) +  (n_2^a - \tau
 n_2^b)(\xi_2 x^+ + a_2) \vert^2 \Biggr\} }
diverges.  This leads to potential divergences in one-loop
amplitudes. Similarly, there are potential divergences  in
propagators and vev's of composite operators.   While such
divergences are perhaps cured by supersymmetric cancellations, we
will see below that the NCNC's lead to singularities in
wavefunctions analogous to those of the parabolic orbifold.

One can show that for $r\geq 3$ generators the generic model has
no CNC's or fixed points but suffers from NCNC's. The difference
{}from the $r=2$ case is that now this happens for all $x^+$,
since for all $x^+$ there will be infinitely many Diophantine
approximations $n_1\omega_1 + n_2 \omega_2 + n_3 \omega_3 \cong
0$.

In view of these remarks we are motivated to impose criterion 6
above. It then follows from the classification of the two
generator models above that:

\noindent {\it The only time dependent orbifolds of $\IR^{1,3}$
with nonsingular physics are the null branes with $R>0$. }

\subsec{Comments on the orbifold groups for $d>2$}

It would be very interesting to extend the  analysis of the
previous section to $9+1$ dimensions. It is possibly useful in
this context to represent Lorentzian spacetime as $2\times 2$
Hermitian matrices over the octonions ${\bf O}$ since one can
then identify the Lorentz group with $SL(2,{\bf O})$
(suitably-defined) \manogue. However we will follow another
approach using the results of \figueroai. We will only make some
general comments and will not achieve a full classification of
the possibilities.

Using the results of Appendix A of \figueroai\ is straightforward
to show that the criterion of some unbroken supersymmetry and
time dependence implies that the orbifold group $\Gamma \subset
\CP(1,9)$ must be a discrete subgroup of
\eqn\tendeedg{
G=
\bigl( {\rm Spin}(7) \sdtimes \IR^8 \bigr) \sdtimes \IR^{1,9}
}
To describe the action of $G$ as a group of Poincar\'e
transformations we denote vectors by $(x^+, \vec x, x^-)\in
\IR^{1,9}$ with $\vec x \in \IR^8$. The general group element
then acts by
\eqn\gengee{ \eqalign{ x^+ & \to x^+ + R^+ \cr \vec x  & \to
\sigma\cdot ( \vec x + \vec v x^+  + \vec R ) \cr x^- & \to x^- +
\vec v\cdot \vec x + \half \vec v^2 x^+ + R^- \cr} }
where $R^\pm$ are real, $\vec v, \vec R \in \IR^8$, and $\sigma$
is a rotation in the $ {\rm Spin(7)}$ subgroup of the $SO(8)$
isometry group of $\IR^8$ fixing the covariantly constant spinor.

A natural subgroup to consider is that generated by the
transformations \gengee\ with  $R^+=0$. A simple computation of
$(x-\Lambda\cdot x)^2$ shows that identifications by elements of
this subgroups do not generate CTC's, and in this sense the
subgroup defined by $R^+=0$ may be considered an analog of the
group $\CS$ of section 7.2. It is generated by ${\rm Spin}(7) $
rotations together with translations  in $x^-$ by $R^-$, denoted
$t(R^-)$, and the transformations $\Lambda(\vec v, \vec R)$
defined by \gengee\ with $R^+=R^-=0$ and $\sigma =1 $. The
subgroup of elements $t(R^-) \Lambda(\vec v, \vec R)$ which is
$\IR^8 \sdtimes (\IR^8  \oplus \IR)$ forms a 17-dimensional
Heisenberg group generalizing the 5-dimensional Heisenberg group
we found before. Indeed: $t(R^-)$ is central in $G$ and
\eqn\heisenagain{ \Lambda(\vec v_1, \vec R_1) \Lambda(\vec v_2,
\vec R_2) = t(\vec v_1 \cdot \vec R_2) \Lambda(\vec v_1 + \vec
v_2, \vec R_1 + \vec R_2) }
There is an analogous Heisenberg group acting on $\IR^{1,d+1}$
for any $d$.

The presence of the ${\rm Spin}(7)$ factor in $G$ is a new
feature of higher dimensions, not seen in the 3+1 dimensional
null brane discussed above. The inclusion of rotations in the
classification of models is a very  interesting subject to which
we hope to return. One important point is that the rotations in
$Spin(7)$ cannot be  arbitrary, even when accompanied by a shift
$\vec R$.  This is implied by the convergence of the 1-loop
amplitude. The basic point can already be made in the  standard
Melvin model $(\IC \times \IR)/\IZ$, defined by the orbifold
action $(z,y) \to  (e^{2\pi i \gamma} z, y+R) $ where $z$ is
complex, $y$ is real, and $\gamma$ is a real rotation angle. The
convergence of the 1-loop partition function for a fixed $\tau$
is controlled by  the convergence of
\eqn\effectivesum{ \sum_{s\in \IZ}  e^{- \beta  s^2} {1\over
\parallel s \gamma \parallel^2} }
where $\parallel s \gamma \parallel$ is the distance to the
nearest integer.  The denominator in \effectivesum\ arises from
the entropy of states in the $s$ twisted sector.  If this entropy
grows with $s$ sufficiently fast, the sum \effectivesum\ can
diverge. In number theory it is shown that there are some
irrational numbers $\gamma$ such that there can be a series of
extremely good rational approximations to $\gamma$.  For these
$\gamma$'s the denominator term in \effectivesum\ can overwhelm
the Gaussian falloff term in \effectivesum.  An example of such an
irrational number $\gamma$ can be constructed as follows.  Let
$f(n)$ be a very rapidly growing function, for example $f(n+1) =
A^{f(n)}$ for a large constant $A$, and set
\eqn\exmple{ \gamma = \sum_{n=1}^{\infty} {1\over 10^{f(n)} } }
Then, equation \effectivesum\ diverges. In the remainder of this
paper we consider only quotients by subgroups of the Heisenberg
group.

Turning our attention now to the causal structure of the quotient
spaces $\CF_{x^+} = \IR^{d+1}/\Gamma$ we note that there   will
be no NCNC's if the generating vectors $\vec v_i x^+ + \vec R_i$
of the torus in $\IR^d$ are linearly independent (over $\IR$) for
all $x^+$. Put differently, if we assemble these vectors into an
$d\times r$ matrix $v x^+ + R$, we require that this matrix have
constant rank for all $x^+$.

\subsec{Wavefunctions on the generalized orbifolds}

Now let us discuss the analog of the $U$-functions for general
orbifolds associated with subgroups of the Heisenberg group
discussed above.  The general solution of the wave equation with
fixed $p^+\not=0$ in $\IR^{1,d+1}$ can be written as:
\eqn\sumplwv{ U_{\chi,p^+}(X) := \int_{\IR^d}  d \vec p~
\chi(\vec p)~ e^{i (P,X)} }
Here we write $X= (x^+, \vec x, x^-)$ with $\vec x \in \IR^d$ and
similarly \eqn\pvecon{ P = \bigl( p^+, \vec p , p^-= {\vec p^2 +
m^2\over 2p^+} \bigr) , } while $(P,X) = - p^+ x^- - p^- x^+ +
\vec p \cdot \vec x$ is the Lorentz invariant inner product. It
is easy to show that
\eqn\plwvii{
U_{\chi,p^+}(\Lambda(\vec v, \vec R)\cdot X)=
\int_{\IR^d}  d \vec p ~\chi(\vec p + \vec v p^+ ) ~ e^{i (\vec
p + p^+ \vec v)\cdot
\vec R} ~ e^{i (P,X)}
}

Now let us consider wavefunctions on an orbifold with $r$
generators $g_i = \Lambda(\vec v_i, \vec R_i), i=1,\dots, r$.
{}From \plwvii\ we learn that invariant wavefunctions must satisfy
\eqn\plwviii{
 \chi(\vec p + \vec v_i p^+ ) =  e^{-i (\vec p + p^+ \vec v_i)\cdot
\vec R_i } \chi(\vec p) \qquad i=1,\dots, r
}

Let us now consider the two sources of difficulty, identified
below equation \psili, in the amplitudes. First we examine the
condition of nonsingularity of the wavefunction for all $x^+$. In
order to analyze it, we need to find a more explicit solution to
\plwviii. Let $H_v\subset \IR^d$ be the linear span of the
vectors $\vec v_i$. Suppose we can find a quadratic form $Q$ on
$\IR^d$ and a vector $\vec b\in \IR^d$ such that
 \eqn\plwvv{ \eqalign{
 p^+ Q\vec v_i & = \vec R_i \qquad i = 1,\dots r \cr
 \vec b \cdot \vec v_i &  = \half \vec v_i \cdot \vec R_i
 \qquad i=1,\dots, r \cr} }
Then a solution to equation \plwviii\ can be written as
\eqn\plwviv{
\chi(\vec p) = e^{-i\half \vec p Q \vec p - i \vec b\cdot \vec p}
\tilde\chi(\vec p)
}
with
\eqn\plwvvi{
\tilde \chi(\vec p + p^+ \vec v ) = \tilde \chi(\vec p)
}
i.e.\ $\tilde \chi$ is a periodic function along the hyperplane
$H_v$ spanned by $\vec v_i$ with periods $\ p^+\vec v_i$, and can
be written as a sum of plane waves.

At this point, we sacrifice some generality and assume that $\vec
v_i$ are linearly independent and therefore $H_v$ has dimension $r$.
Then it is easy to find $Q,\vec b$ as follows.  Construct $\vec
w_i, i=1,\dots, r$ and $\vec u_{i'}, i'=1,\dots d-r$ such that
\eqn\systemvc{
\eqalign{
w_{aj} v_{ai} & = \delta_{ij} \cr
u_{aj'} v_{ai} & = 0 \cr
u_{ai'} u_{aj'} & = \delta_{i'j'}\cr}
}
Here and below, repeated indices are summed.  We can now decompose
$\vec R_i$ in the basis $\vec w_i, \vec u_{i'}$:
\eqn\arrvect{
R_{ai} =  w_{aj} I_{ji} +   u_{ai'}
N_{i' i}
}
Note that, by the condition that $\Gamma$ be Abelian, $I_{ji} =
\vec v_j \cdot \vec R_i $ is {\it symmetric} so  we can set
\eqn\solvequee{
p^+ Q_{ab} = w_{ai} I_{ij} w_{bj} + u_{ai'} N_{i'i} w_{bi} +
u_{bi'} N_{i'i} w_{ai}
}
while $\vec b = \sum_{i=1}^r \half (\vec v_i \cdot \vec R_i )
\vec w_i $ will do.

Now let us ask if the wavefunctions $U_{\chi,p^+}$ become
singular for any $x^+$. We introduce coordinates in momentum
space appropriate to the decomposition $\IR^d = H_v \oplus
H_v^\perp$:
\eqn\momcoord{
p_a =  v_{ai} p_i +   u_{ai'} q_{i'}
}
Then from \plwvvi\  we can Fourier decompose and write:
\eqn\plwvvii{
\tilde \chi(p_i,q_{i'} ) = \sum_{\vec n} \tilde \chi_{\vec n}
(q_{i'}) e^{2\pi i n_i p_i/p^+}
}
and $\tilde \chi_{\vec n}(q_{i'})$ can be functions of rapid
decrease. Let us similarly write
\eqn\decexx{
X_a = w_{ai} y_i + u_{ai'} z_{i'}
}
Then, up to a constant, the wavefunction $U_{\chi,p^+}$ can be
written:
\eqn\wavefunc{
\eqalign{
&
e^{-ip^+ x^-} \sum_{\vec n}
\int \prod_{i'=1}^{d-r} dq_{i'} \tilde \chi_{\vec n}(q_{i'})
e^{-{i\over 2p^+}(q_{i'} q_{i'} + m^2)}
e^{i z_{i'} q_{i'} } \cr
&
\int \prod_{i=1}^r dp_i e^{-{i\over 2p^+} p_i p_j (I_{ij} +
x^+ v_i\cdot v_j)}
e^{-{i\over p^+} q_{i'} N_{i'i}p_i}
e^{i p_i\bigl( y_i - \half (\vec v_i \cdot \vec R_i) + 2
\pi n_i/p^+\bigr)} \cr}}

Now we see that the condition of nonsingularity of the
wavefunction for all $x^+$ fits in very nicely with the condition
for the absence of NCNC's. Recall that the condition for the
absence of NCNC's is that the $d\times r$ matrix
\eqn\noncncs{
v_{ai} x^+ + R_{ai}
}
have constant rank for all $x^+$. Suppose this condition is
satisfied. Then,  there is no possible singularity in
$U_{\chi,p^+}$ unless the quadratic form $I_{ij} + x^+ \vec v_i
\cdot \vec v_j$ has a nonzero eigenvector $p^*_i$ of eigenvalue
zero. By our assumption on the rank of $v_{ai}$ this will only
happen at special critical times  $x^+_c$. In this case   the
Gaussian integral on $p_i$ degenerates to a delta function.
However, as long as this vector $p^*_i$ is not also in the kernel
of $N_{i'i}$ the argument of the delta function depends
nontrivially on the  $q_i'$ and the subsequent $q_{i'}$ integral
is   nonsingular. But the condition that any null vector of
$I_{ij} + x^+ \vec v_i \cdot \vec v_j$ is not in the kernel of
$N_{i'i}$ is precisely the condition that \noncncs\ have constant
rank!

{\it Thus, we conclude that the condition of the nonsingularity
of the wavefunctions is precisely the condition for the absence
of nearly closed null curves in the quotient $\CF_{x^+} =
\IR^{d+1}/\Gamma$ for all $x^+$.  }

Let us finally comment briefly on the suppression of the large
energy component. Evidently, since $\chi$ in \sumplwv\ is a
(quasi) periodic function along $H_v$, it cannot have rapid
decrease in these directions. It can have rapid decrease along
the orthogonal space $H_v^\perp$. For example, in \plwvvii\ we
can take $\tilde \chi_{\vec{n}} (q_{i'} )$ to be functions of
$q_{i'}$ of rapid decrease. Then, as in our discussion in the null
brane case in sec. 3, it is possible to suppress large momenta
along  $H_v$ provided that $N_{i'i}$ satisfies certain properties.
More explicitly, from \wavefunc\ the integration over $q_{i'}$
with  $\tilde \chi_{\vec{n}} (q_{i'} )$ will generate a rapidly
decreasing function of $\sum_i N_{i'i} p_i$ for fixed $z_i$. Thus
if $\sum_i N_{i'i} p_i$ span\foot{This requires that $r \leq d-r$
and the matrix $N_{i'i}$ should have rank $r$.} $H_v$ then large
momenta in all directions including $H_v$ will be suppressed.

\bigskip
\centerline{\bf Acknowledgements}

We thank S.~Sethi for stressing to us the importance of the
null-brane. We also thank G.~Horowitz, J.~Polchinski and
E.~Silverstein for encouraging us to examine the null-brane. We
thank C.~Bachas, T.~Banks, J.~Froehlich, G.~Horowitz, C.~Hull,
J.~Maldacena, J.~Polchinski, S.~Shenker, E.~Silverstein,
L.~Susskind, and E.~Witten for discussions. HL and GM were
supported in part by DOE grant \#DE-FG02-96ER40949 to Rutgers. NS
was supported in part by DOE grant \#DE-FG02-90ER40542 to IAS. GM
would like to thank L. Baulieu and B. Pioline  for hospitality at
the LPTHE and the Isaac Newton Institute for hospitality during
the completion of this work.

\listrefs

\bye